\documentclass[12pt,fleqn,reqno]{amsart} %[12pt,reqno]{amsart} %[12pt,twoside]

\usepackage{amsmath,euscript}
\usepackage{amsthm}
\usepackage{amssymb}
\usepackage{graphicx}
\usepackage{mathrsfs}
\usepackage{epsf}
\usepackage{psfrag}
\usepackage{enumerate}
\usepackage{amsfonts,amssymb,amsthm,amsmath,latexsym,marginnote,color}

\usepackage{hyperref}

\usepackage[numbers]{natbib}
\usepackage{enumitem} % for labeling in enumeration
\usepackage{xspace}
\usepackage[margin=1.3in]{geometry}
\usepackage{soul}

\newtheorem{thm}{Theorem}

\newtheorem{prop}[thm]{Proposition}
\newtheorem{cor}[thm]{Corollary}

\theoremstyle{remark}

\theoremstyle{definition}

\numberwithin{thm}{section}
\numberwithin{equation}{section}
\newcommand{\nc}{\newcommand}

\nc{\la}{\label}
\nc{\ba}{\begin{array}}
\nc{\ea}{\end{array}}
\nc{\bs}{\begin{split}}
\nc{\es}{\end{split}}

\newcommand{\R}{\mathbb{R}}
\newcommand{\C}{\mathbb{C}}
\newcommand{\Z}{\mathbb{Z}}

\newcommand{\bT}{\mathbb{T}}

\newcommand{\bH}{\mathbb{H}}

\newcommand{\cA}{\mathcal{A}}

\newcommand{\cC}{\mathcal{C}}
\newcommand{\cE}{\mathcal{E}}
\newcommand{\cF}{\mathcal{F}}
\newcommand{\cH}{\mathcal{H}}

\newcommand{\cL}{\mathcal{L}}
\nc{\al}{\alpha}
\nc{\del}{\delta}
\nc{\h}{\delta}
\nc{\G}{\Gamma}
\nc{\et}{\eta} 
\nc{\g}{\gamma}
\nc{\gam}{\gamma}
\nc{\ka}{\kappa}
\nc{\lam}{\lambda}
\nc{\Lam}{\Lambda}
\nc{\Om}{\Omega}
\nc{\om}{\omega}

\nc{\ta}{\tau}
\nc{\w}{\omega}
\nc{\io}{\iota}
\nc{\z}{\zeta}
\nc{\s}{\sigma}
\nc{\Si}{\Sigma}
\nc{\vphi}{\varphi}

\nc{\e}{\epsilon}

\newcommand{\fh}{\mathfrak{h}}

\nc{\bP}{\bar{P}}
\nc{\bQ}{\bar{Q}}
\nc{\x}{\underline{x}}
\nc{\y}{\underline{y}}

\nc{\ran}{\rangle}
\nc{\lan}{\langle}
\def\<{\langle}
\def\>{\rangle}

\newcommand{\ra}{\rightarrow}

\newcommand{\ls}{\lesssim}

\newcommand{\one}{\mathbf{1}}

\newcommand{\Ran}{\operatorname{Ran}}
\newcommand{\Null}{\operatorname{Null}}

\renewcommand{\Re}{\mathrm{Re}} % Real part
\renewcommand{\Im}{\mathrm{Im}} % Imaginary part

\newcommand{\im}{{\rm Im}}
\newcommand{\Tr}{\mathrm{Tr}}
\newcommand{\tr}{\mathrm{Tr}}

\nc{\bfone}{{\bf 1}}
\nc{\1}{{\bf 1}}

\newcommand{\p}{\partial}

\newcommand{\n}{\nabla}

\newcommand{\curl}{\operatorname{curl}}
\newcommand{\CURL}{\operatorname{curl}}
\newcommand{\divv}{\operatorname{div}}

\newcommand{\na}{\nabla_a}

\newcommand{\DETAILS}[1]{}

\newcommand{\hf}{H_f}
\renewcommand{\part}{{\rm part}}
\nc{\den}{\text{den}}
\nc{\ex}{\text{xc}}
\nc{\Ex}{\text{Xc}}

\nc{\jx}{\langle x \rangle}

\def\qf{\chi} %{\varphi}

\newcommand{\lat}{\mathcal{L}} %{\lambda}}
\newcommand{\LAT}{\mathcal{L}} %{\mathcal{\lambda}}

\newcommand{\tor}{\bT}

\definecolor{green}{rgb}{0.0, 0.5, 0.5}

\definecolor{lgray}{gray}{0.9}
\definecolor{llgray}{gray}{0.95}
\definecolor{lllgray}{gray}{0.975}

\newcommand{\blue}{\color{blue}}

\usepackage{soul}
\setstcolor{red}

\begin{document}

\title[On Differential Equations of Quantum Mechanics]{Differential Equations of Quantum Mechanics}

\author{I. M. Sigal} %\footnote{The research on this paper is supported in part by NSERC Grant No. NA7901.}  
\address{Israel Michael Sigal, Department of Mathematics, University of Toronto, %40 St. George street, 
 Toronto,   M5S 2E4, Canada}
\email{im.sigal@utoronto.ca}

%\subjclass[2000]{Primary 35Q55 ; 35B34}

%\keywords{Quantum propagation, Maximal velocity bounds, Schr\"odinger equation, Hartree equation}

\begin{abstract} We review very briefly  the main mathematical structures and results in some important areas of Quantum Mechanics involving PDEs  and  formulate open problems.\end{abstract}

\maketitle

%\markboth{}{}

%\tableofcontents

\date{January 9, 2022}

\tableofcontents

%\bigskip
\section{Preface} %{Introduction} 
  This note follows closely 
  the talk I gave at the ONEPAS and then repeated in a modified form at the MCQM\footnote{ONEPAS and MCQM stand for Online Northeast PDE and Analysis Seminar and Mathematical Challenges in Quantum Mechanics.} seminar  series. I inserted a few details and explanations in the text (often as footnotes), expanded the comments on the literature at the end of the main text along with three appendices giving some precise definitions omitted in the talk (and the main text).   
  My goal was to describe the main mathematical structures and results of an important area of Quantum Mechanics involving PDEs  and to formulate open problems. 

Some of the open problems are closely related to the mainstream PDEs; others would draw blank for most of the PDE practitioners. 
The problems of the first type 
deal with quantum fluids such as superconductors, superfluids and Bose-Einstein condensates which are natural to describe in terms of 
 operators (density operators, etc), rather than functions, though passing to integral kernels one could  produce the standard PDE  description.

 The problems of the  second type concern particle systems interacting with quantized radiation, i.e. photons. The key goal   here is to describe the processes 
   of emission and absorption of radiation, say light, by     systems of matter such as atoms and molecules. These are, obviously, not very fanciful problems and they 
 deserve close attention. 

  Naturally, the material selected in this article adheres closely to my own research and is not comprehensive whatever interpretation of this word is used. 

All references to the literature are collected in Section \ref{sec:liter}. I tried to be fair and acknowledge all recent contributions to the subjects covered.  
 However, I am sure I missed many worthy works and I will appreciate any information about those.

%%%%%%%%%%%%%%%%%%%%%
\subsection*{Acknowledgments}
I am grateful to S\'ebastien  Breteaux, Thomas Chen, Ilias Chenn, J\'er\'emy Faupin, Zhou Gang, Marcel Griesemer, Stephen Gustafson Lars Jonsson, Tim Tzaneteas and, especially, Volker Bach,  J\"urg Fr\"ohlich and Avy Soffer for enjoyable collaboration and to Rupert Frank, Gian Michele Graf, Christian Hainzl, for stimulating discussions on the topics touched upon in this review.

It is a pleasure to thank Javier Gomez-Serrano, Benoit Pausader, Fabio Pusateri, Ian Tice and Claudio Cacciapuoti, Raffaele Carlone, Michele Correggi for the invitations to speak in their seminar series and to Javier, Benoit, Fabio and Ian for suggesting to write a review and  encouragement, and  Benoit, for many constructive remarks. 

 The author's  research 
   is supported in part by NSERC Grant No. NA7901.

%%%%%%%%%%%%%%%%%%%%%%%%%%
\section{Schr\"odinger equation} 

Quantum systems of $n$ particles in the space $\R^3$ are described by the {\it Schr\"odinger equation}\footnote{The relation of solutions of the Schr\"odinger equation  
 to quantum observable effects   is  %\st{dealt with by} 
 a subject matter of  Quantum Mechanics.}

\begin{equation*}i \p_t \Psi_t=H_n \Psi_t, \tag{SE}\end{equation*} 
 where $\Psi_t=\Psi_t(x_1, \dots, x_n)$, a family of Sobolev functions of the particle coordinates $x_1, \dots, x_n$ in $\R^3$ and $H_n $ is an $n-$particle Schr\"odinger operator. For  
$n$ particles of masses $m_1, \dots, m_n$ interacting via $2$-body potentials $v_{ij}$ and with external potentials $V_i$, $H_n $ has the form
\begin{equation}\label{Hn}H_n:=\sum_1^n h_{x_i} +\sum_{i<j} v_{ij}(x_i-x_j),\end{equation}
where $h_{x_i}=\frac{-1}{2m_{i}}\Delta_{x_i}+V_i(x_i)$ is a one-particle Schr\"odinger oprerator with an external potential $V_i(x)$ acting on the $i$-th particle variable $x_i$.  
The function $\Psi_t$, called the wave function, gives probability distributions at time $t$ for various physical observables.  
  
  \medskip
  
One of the first results going back to J. von Neumann and elaborated  
 by B. Simon    can be formulated as 
\[\text{ global existence $\Longleftrightarrow$  self-adjointness of }H_n.\]

  \medskip

By the Kato-Rellich theorem, the operator $H_n$ is self-adjoint for a fairly large class of potentials.  Hence, the mathematical task, 
 as suggested by the underlying physics, 
 is to \[\text{ describe the space-time behaviour of solutions of (SE). }\]

  \smallskip

\noindent The {\it main dichotomy} here can be formulated as
 \[\text{ stability vs. collapse or disintegration}.\]

  \medskip

By  stability we mean the {\it localization in space} and periodicity in time. This can be further enlarged on %elaborated 
 and reduced to spectral properties of $H_n$: 

  \medskip
  
\begin{itemize} \item stability w. r. to collapse ($\inf H_n %\ge -C
>-\infty$) 
\item stability w. r. to break-up %(gap($\inf H$, rest)$>0$).
($\inf H$ is an eigenvalue).\footnote{The condition $H_n\ge -C>-\infty$, which follows from a refined uncertainty relation, says that the particles are not sucked into attractive Coulomb singularities, contrary to the prediction of classical physics. The condition gap($\inf H$, rest)$>0$ implies that, for initial conditions close to the ground state (eigenfunctions corresponding to the lowest energy $\inf H$),   
 the solution to (SE) 
remains concentrated in a bounded set in the configuration space $\R^{3n}$ (with the motion of the centre of mass factored out)   
and that this property is stable w.r. to perturbations of the Hamiltonian.} 
\end{itemize}
Thus, the stability explains the existence of structures of matter such as atoms, molecules,
 ..., stars.

One can refine the first type of stability as the extensivity of the energy, i.e. $H_n\ge -c n.$
The latter is known as the stability of matter.

  \medskip
  
We say  a system undergoes decay, or, more precisely, local decay, if the probability ($\int_Q|\Psi_t|^2$) that it occupies any bounded domain ($Q\subset \R^3$) in the physical space vanishes with time. The description of all possible scenarios of such an evolution is provided by   the scattering theory.

 \bigskip

%----------------------- 5 -------------------
\paragraph{\bf Scattering.} The main mathematical problem of the scattering theory -- the {\it asymptotic completeness} -- states that

\bigskip

As time progresses, a quantum system settles in a superposition of states %in each of which it is broken into
consisting of collections of  stable freely moving fragments.

%--------------------------- 15 picture -----------------------------

%$$\text{ Figure }$$

\begin{thm}[Asymptotic completeness]\label{thm:AC} Suppose that the pair potentials $v_{ij}(x_i-x_j)$ entering $H_n$ satisfy $v_{ij}(y)=O(|y|^{-\mu})$, as $|x|\ra \infty$, with $\mu>\sqrt3-1$. Then the  asymptotic completeness holds.
\end{thm}

\bigskip

\noindent {\it Open problem}: Prove the  asymptotic completeness for $v_{ij}(y)=O(|y|^{-\mu})$, with $\mu\le \sqrt3-1$.

%----------------------- 5 -------------------
\section{Including photons (Nonrelativistic QED)}\label{sec:QED}

To describe the real (or at least visible) world, we have to couple  the particles to {\it photons} (i.e to {\it quantized electromagnetic field}). 
The dynamics of the resulting system is  described again by
 the Schr\"odinger equation,\footnote{This  Schr\"odinger equation was used by P.A. Dirac and E. Fermi already in the early days of Quantum Mechanics, see Fermi's review \cite{Fermi}.}
\begin{equation}\label{SEqed}
 i \p_t\Psi_t =H_\kappa\Psi_t,
 \end{equation}
 but with the more complex Hamiltonian  $H_\kappa$ acting on the state space 
${\cH}:=\cH_p\otimes \cH_f$, which is the tensor product of the spaces of particles and photons. This Hamiltonian is given by  
\begin{equation}\label{Hqed}
H_\kappa = \sum_{j=1}^n \frac{1}{2 m_j} \big( - i \nabla_{x_j} - \kappa A_\xi ( x_j ) \big)^2 + U (x) + H_f .
\end{equation}
 Here,  
  $\kappa$ is the  
 particle charge,    $U(x)$, $x = ( x_1 , \ldots , x_n )$, is the 
   total potential effecting the particles,\\ \medskip 
  \qquad     $A_\xi =\xi * A$ is the 
       UV-regularized, {\it quantized vector 
     potential} $ A$ and\\ \medskip
   \qquad   $H_{f}$ is the 
  photon Hamiltonian.\footnote{For the definitions of $A$ and $H_f$, see Appendix \ref{sec:QED-ham}.}

\medskip

The key phenomena one would like to describe are  
  emission and absorption of the electromagnetic radiation. Physical description of these processes translates into the following mathematical problems:  

- the existence of the ground state 

- the instability of excited states 

- the emergence of resonances.  

The ground and excited states are eigenstates of $H_\kappa$ with the smallest and remaining 
   eigenvalues (energies), respectively. The  resonances are thought of physically as  `metastable' states, or `bound states with finite life-times'. Mathematically, they correspond to complex poles of an analytic continuation of the resolvent of $H_{\kappa}$ across the continuous spectrum to the second Riemann sheet. 
\footnote{The instability of excited states means that $H_\kappa$ has no eigenvalues in a small neighbourhood of the excited eigenvalues  of $H_{\kappa=0}$. As $\kappa$ is `turned on', the real poles of the resolvent of $H_{\kappa=0}$ corresponding to the excited eigenvalues  migrate to the second Riemann sheet.

To connect this instability to, say,  the spontaneous emission of photons, one shows that the solution starting with the initial condition which is an excited eigenvector  of $H_{\kappa=0}$, i.e. the tensor product of  an excited eigenvector  of the particle system and the photon vacuum,  
 describes the particle system descending 
  to the ground state, with the difference in energy carried out by a departing photon(s).}

Establishing the properties  above and giving estimates of the renormalized  
 energies and life-times are the main tasks of mathematical theory of radiation.

\begin{thm}[Problem of radiation]\label{thm:rad} Assume $U(x)$ is a Kato-type potential and the Fermi Golden rule holds. Then for $\kappa>0$ sufficiently small,

(a) $H_\kappa$  has a (unique) ground state exponentially localized in the particle coordinates, 

(b) $H_\kappa$  has no excited states outside of a sufficiently small neighbourhood of the continuous spectrum, and 

(c) $H_\kappa$  has resonances of the multiplicities equal to those of the vanished eigenvalues and which converge to the latter as $\kappa\ra 0$. 
\end{thm}

The Fermi Golden rule expresses the effectiveness of the coupling of the particles to the electromagnetic field (that there is no accidental decoupling). The uniqueness of the ground state depends on symmetries present and could fail. Statements (b) and (c) express the instability of excited states and their turning to the resonances.

We discuss briefly difficulties arising in proving the theorem above. Since the photons are massless, they can be born out of the vacuum and absorbed back to the latter, as well as emitted and absorbed by the particle systems, in arbitrary large numbers and form dense fluctuating clouds around the particles. This leads to divergences 
in the formal perturbation series  for various physical quantities (e.g. the energies of the ground states and resonances), the phenomenon  known as the  {\it infrared problem}.\footnote{The photon clouds lead to renormalizations of various physical quantities (such as the mass of electron and energies of the ground state and resonances), which would diverge if the ultra-violet cut-off (UV) is removed.}  

If we think of $H_\kappa$  as a perturbation of  the Hamiltonian $H_{\kappa=0}$ of the decoupled system in which the particles and photons do not interact, 
then we run into another,  related, manifestation of the infrared problem: 
 the bound state energies of the Hamiltonian $H_{\kappa=0}$ 
  are not isolated.\footnote{Indeed,  the bound state energies of $H_{\kappa=0}$ are exactly   the bound state energies, $E^{(k)}_{\rm part}$, of the particle Hamiltonian  \[H_{\rm part} = \sum_{j=1}^n \frac{1}{2 m_j} \big( - i \Delta_{x_j}  \big) + U (x).\] However, the values $E^{(k)}_{\rm part}+\lam, $ where $\lam\in \s(H_f)$ is an arbitrary photon energy, 
 also belong to the spectrum of $H_{\kappa=0}$.  Since the photon energies fill in the semi-axis $[0, \infty)$, for each bound state energy $E^{(k)}_{\rm part}$, 
 there is a continuous spectrum branch $[E^{(k)}_{\rm part}, \infty)$ of $H_{\kappa=0}$.   
}

 The standard perturbation theory fails in treating such eigenvalues and one needs a {\it new} theory. The {\it spectral renormalization-group theory} developed to prove Theorem \ref{thm:rad} 
  is exactly the theory which deals with this problem.

The next problem is to show that, as we experience in everyday life, most of the photons born out by a particle system escape it at some characteristic time and travel freely unless they
encounter with another particle system. The mathematical formulation of this property leads to the problem of asymptotic completeness which we have already met in the previous section. 

To formulate the result here, let $N_{\rm ph}$ be the quantum observable (self-adjoint operator) of the number of  photons, $\Sigma:=\inf\s_{\rm ess}(H_{\rm part})$, the ionization threshold,  and $\Psi_t$, a solution to \eqref{SEqed}. Then, we have

\begin{thm}[Rayleigh scattering]\label{thm:Rayl} 
 Assume that initially the energy is localized below the ionization threshold, $\Sigma,$  of the particle system and 
that $\sup_t\lan \Psi_t, N_{\rm ph} \Psi_t \ran <\infty$ 
  (satisfied in special cases). Then the asymptotic completeness holds.
\end{thm}

\bigskip
\noindent {\it Open problem}: Prove that $\sup_t\lan \Psi_t, N_{\rm ph} \Psi_t \ran <\infty$ for general particle systems (like atoms).

\section{Effective  Equations} \label{sec:eff-eqs} 
 To describe systems of a large number of particles, say from $10$ to $10^{20}$, it is necessary %/imperative (?)
 to design effective approaches giving some aggregate, or `collective', information. % about a system in question. 
Often, such approaches provide information which is practically impossible to extract from solving the original equations even if this was feasible. In the quantum many-body problem, the key effective approach is the Hartree-Fock approximation, trading the number of particles for the nonlinearity. This approximation, its natural extension and the effective equations it leads to are described %this approach
 below.
 
 \bigskip
 
 \paragraph{\bf Hartree and Hartree-Fock equations and their extensions.} Consider  a system of $n$ identical {\it bosons} or {\it fermions} whose evolution is described by  
 the Schr\"odinger equation \begin{equation*}i \p_t \Psi_t=H_n \Psi_t, \tag{SE}\end{equation*} 
where $H_n$ is the Schr\"odinger operator  given in \eqref{Hn}, with $m_{i}=m$, $v_{ij}=v$ and $V_{j}=V$. 
To obtain an effective, one-particle approximation for large $n$, we restrict (SE) to the {\it Hartree} and  {\it Hartree-Fock} states given by symmetric and anti-symmetric products of one-particle wave functions:
\begin{equation}\otimes_1^n \psi\ \quad \text{ and }\  \quad \wedge_1^n \psi_i,\end{equation} 
 This leads to equations for $\psi$ and $\psi_1, \dots, \psi_n$ - the  {\it Hartree} (H)  and    {\it Hartree-Fock} (HF) equations, 
widely used in physics and chemistry.

\bigskip

However, these equations fail to  account for quantum fluids i.e. quantum gases exhibiting some quantum behaviour at the macroscopic scale such as superconductivity,  superfluidity and Bose-Einstein (BE) condensation. %ors, superfluids and Bose-Einstein (BE) condensates. 
For this, one needs {\it another conceptual step}.

\bigskip

\paragraph{\bf Non-Abelian random Gaussian fields and Wick states.} 
In rigorous quantum statistical mechanics, the states are described by linear, positive (normalized) functionals, $\om$, on a $C^*$-algebra of observables, $\cA$ and their evolution  is  given by the 
von~Neumann-Landau equation 
\begin{align}\label{vN-eq} 
     \partial_t\om_t(A) =- \om_t(i[H, A]) \,,\ \forall A\in   \cA,  
\end{align}
 where $\om_t$ is the state at time $t$.

The simplest and in a sense a single most important class of (fixed-time) states, $\chi$, consists of quantum (non-Abelian) generalization of random Gaussian fields, i.e. states  
 uniquely determined by the {\it two-point correlations}\footnote{For a moment, we ignore the expectation $\chi (\psi(x))$.}: 
\begin{equation} \label{psi*psi-correl}\
     \chi (\psi^{*}(y ) \,  \psi(x)), \end{equation} 
where $\psi(x)$ and $\psi^{*}(x )$ are quantum fields adjoint to each other (the annihilation and creation operators). 
These are exactly the {\it HF states}: If 
 $\chi_t$ is an evolving HF state, then the {\it operator $\g$ with the integral kernel} $\chi_t(\psi^{*}(y ) \,  \psi(x))$ 
   satisfies the {\it HF equation}  
\begin{align}\label{HF-eq}   
   &  i\partial_{t}\gamma =[h_{\gamma},\gamma],
\end{align} where  $h_{\gamma}$ is the $\g$-dependent, one-particle Schr\"odinger operator,  %given by  
 \begin{align}\label{hgam}
& h_{\g}  =-\Delta+V  
+ v * \rho_\gamma - v^\sharp \g, \end{align}
with $\rho_\gamma(x, t):= \gamma(x, x, t)$ (=\ --charge density) and  $v^\sharp \g $ the operator with the integral kernel $(v^\sharp \g)(x, y) = v(x - y)\g (x;y)$. The terms $v * \rho_\gamma$ and $v^\sharp\, \gamma$ are the direct (`electrostatic') and exchange (HF) self-interaction energies.

However, the above states are not the most general `quadratic' (or quantum Gaussian) states. The most general ones are defined by
 {\it all} one- and two-point correlations\footnote{For more details see \cite{BBCFS} and Appendix E of \cite{CS1}.}
\begin{align}  
 \begin{split}  &    
 \chi_t (\psi(x)), \quad   \chi_t( \psi^{*}(y) \, \psi(x)) \quad \text{and}\ \quad    
   \chi_t(\psi(x) \, \psi(y)). \end{split} \end{align}
 This type of states were introduced by Bardeen-Cooper-Schrieffer  
for fermions and by Bogolubov, for bosons.  
For such states all correlations are either $0$ or are f 
 sums of products of one- and  two-point 
  ones. This is the {\it Wick} property from quantum field theory (QFT)  
 and consequently, we call such states the {\it Wick} states. (In mathematical literature, they are called the quasifree states.)  They give the most general effective one-body description 
  to the $n-$body dynamics. 
 
\medskip

\noindent {\bf Remark.} If we replace the exchange energy operator, $ex(\g):=v^\sharp \g$ in the operator \eqref{hgam} entering the Hartree-Fock equation  \eqref{HF-eq}
 by an multiplication operator by a local function $xc(\rho_\g)$ of $\rho_\g$ 
 which models  {\it exchange} and {\it correlation} energies, we arrive at the time-dependent Kohn-Sham equation, a key equation of the {\it density functional theory} (DFT).

%--------------------------- 1 -----------------------------

 \medskip

\paragraph{\bf Dynamics.}  
 
Restricting the von~Neumann-Landau evolution \eqref{vN-eq} to Wick states 
 yields a system of coupled nonlinear PDE's for  the functions 
\begin{align} \label{phi-def}    
&\phi (x, t) :=     \chi_t( \psi(x)),\\
 \label{gam-def}   &\gamma (x,y, t) :=     \chi_t( \psi^{*}(y) \, \psi(x)),\\  
  \label{al-def}    &\al (x,y, t):=  \chi_t( \psi(x) \, \psi(y)). \end{align}
 These are the (time-dependent)  \textit{Bogolubov-\-de Gennes} (fermions)  and  \textit{Hartree-\-Fock-\-Bogo\-lubov} (bosons) equations\footnote{For more details see Appendix \ref{sec:HFBsyst}.}.

\medskip

For the {\it Bose-Einstein condensation}, 
 $\phi$ is the wave function of the {\it BE condensate} and  $\gamma (x,y, t)$ and $\al (x,y, t)$  
 yield the {\it density  matrix} 
  of the {\it non-condensed atoms} and the {\it `pair wave function'} 
   for the superfluid component. 
   
For {\it fermions}, $\phi (x, t) :=     \qf_t( \psi(x))=0$ and $\g (x,y, t)$ describes the normal electrons, while $\al (x,y, t)$ superconducting ones (more precisely, Cooper pairs of electrons).

In what follows, we associate with the functions $\gamma (x,y, t)$ and $\al (x,y, t)$, the operators $\gamma$ and $ \al$, whose integral kernels these functions are.

As usual for evolution equations, the first mathematical problem here is 

 \smallskip
{\it The well-posedness of the initial value problem.}
   \smallskip

 Physically, the first problem one would like to address is existence of a ground/Gibbs state and its symmetry. 

 \medskip

\paragraph{\bf Ground/Gibbs state and symmetry breaking.}  It is easy to show that the energy, $E(\qf_t):=\qf_t(H)$, if it exists, is conserved. For systems with non-compact symmetries, like translations, the energy is infinite. In this case, one considers `local' or `renormalized' energy. 

The notion of energy allows us to define the key state - the ground 
  state: a {\it static solution minimizing the  (local/renormalized) 
   energy}.  

\bigskip

\noindent  {\it Key problems}: Existence and  symmetry 
  of the ground  state, the excitation spectrum and the dynamics nearby. 
   
\medskip

The most important symmetry is the translational one which holds in the absence of external interactions (potentials). % Not much 
Nothing is known about breaking of this symmetry in the ground states for both equations. 
  
Whenever systems are  considered for positive temperatures, the energy  and the ground state  are replaced by  the free energy and  the Gibbs equilibrium state.

%--------------------------- 1 -----------------------------

\section{Hartree-\-Fock-\-Bogo\-lubov system}
 
We describe the full {\it Hartree-\-Fock-\-Bogo\-lubov (HFB) system} in Appendix \ref{sec:HFBsyst}. Here, we consider the reduced HFB system ($2-$gas model) resulting from neglecting the $\al$-component: 
  \begin{align}
    i\partial_{t}\phi & =     h\phi +v*(|\phi|^{2} +2\rho_{\gamma })\phi,     \tag{GP}\\   
      i\partial_{t}\gamma  & =[h_{\gamma , {\phi} },\gamma ],     \tag{HF} \end{align}
where  
$h=-\Delta+V$ is a one-particle Schr\"odinger operator, $\rho_\gamma(x, t):= \gamma (x;x, t)$, the one-particle density  
  and 
\begin{align}  \notag    &  h_{\gamma , \phi }:=h+v* (\rho_{\gamma }+ |\phi |^{2}). 
\end{align}
These are coupled {\it Gross-Pitaevskii}  and {\it Hartree} equations.  The term  $v* (\rho_{\gamma }+ |\phi |^{2})$ is the direct (`electrostatic') self-interaction produced by the combined charge density $\rho_{\gamma }+ |\phi |^{2}$ of non-condensed and condensed particles (atoms).

\medskip

In addition to the general problems formulated in the previous section,  the following {\it problems}  are of a special interest for the HFB system:

\medskip

-  Bose-Einstein condensation, 

-  Collapse oscillations for $\lam<0$ ({\it correction to the Papanicolaou-Sulem-Sulem collapse law?}).

\medskip

Physically there are two important {\it set-ups} here: 

\medskip

External (attractive or confined) potentials $V$ vs translational invariance ($V=0$).

%{\bf If  $V=0$ and $\int v<\infty$, then the HFB system (the $2-$gas model) has the homogeneous solution $(\phi, \gamma)=(e^{-i\lam t}c, 0)$, with $\lam$ and $c$ satisfying $\lam/|c|^2=\int v $. Thence the question, under which conditions on $v$, this solution is a ground state?}

%--------------------------- 1 -----------------------------

\section{Bogolubov-de Gennes system}\label{sec:BdGeqs}

For {\it fermions}, $\phi (x, t) :=     \qf_t( \psi(x))=0$  and, since the Wick states describe superconductors, 
$(\g, \al)$ are coupled to the {\it electromagnetic field}. With the latter described by the {\it magnetic} potential $a$ 
and the {\it electrostatic potential},  
$\varphi$, and, with the former taken in the Coulomb gauge ($\divv a=0$) and the latter absorbed in the inter-particle (pair) interaction potential, 
 the equations for $\g, \al$ and  
 $a$ read\footnote{For a discussion of gauges, the origin 
   of the BdG equations (in particular, an elimination of the  electric potential) and properties of $\g$ and $\al$  see Appendix \ref{sec:BdGsyst}.} 
     \begin{align} \label{BdG-gam} 
& i\partial_{t}\gamma   =[h_{a, \g},\gamma ]_- + [ v^\sharp \al, \al]_-,\\ 
\label{BdG-al}  & i\partial_{t}\alpha  = [h_{a, \g},\alpha ]_{+} - [ v^\sharp \al, \g ]_{+} + v^\sharp \al,\\ 
 \label{Amp-Maxw-eq}
    - &\partial_t^2 a = \curl^* \curl a + j(\g, a), 
    \end{align} 
 where  $[A,B]_-=A B^* - B A^*$, $[A,B]_+=A\bar B^* + B\bar A^*$, with $\bar A:=\cC A\cC$, with $\cC$ being the complex conjugation,
  $v(x - y)$ is a pair potential, 
$v^\sharp \al $ is, recall, the operator with the integral kernel $(v^\sharp \al)(x, y) = v(x - y)\al (x;y)$,  $j(\g, a)(x) := [-i \na,\g]_+(x, x)$ is the 
   current density, and, finally (cf. Eq. \eqref{hgam}),  
     \begin{align}\label{hgam-a}
& h_{a, \g}  =-\Delta_a+  
 v * \rho_\gamma- v^\sharp \g, 
\end{align}
with $\Delta_a:=(\n + ia)^2$. 
 Here we have assumed, for simplicity, that the external potential is zero, $V=0$, and have chosen the unit electric charge to be $e=-1$.

\medskip

These are the celebrated {\it Bogolubov-de Gennes (BdG) system}. 
 They give the `mean-field' (BCS) theory of superconductivity. 
Eq. \eqref{BdG-gam} is essentially the HF equation coupled to the other two equations and Eq. \eqref{Amp-Maxw-eq}  comes from two Maxwell equations (Amp\`{e}re's and Faraday's laws).  

\bigskip

As was pointed out in Section \ref{sec:eff-eqs}, the key problem here is the {\it existence and symmetry   of the ground  state and the description of the nearby dynamics}.

Experiments show   that at the lowest energy (locally), states of quantum matter typically enjoy the maximal available symmetry. With this in mind, we begin with describing the symmetry group of the BdG system.  

\bigskip

\paragraph{\bf Gauge (magnetic) translational invariance.}  
Arguably,  the simplest and most important symmetry is the translational one. In the magnetic fields this symmetry is broken. However, for constant magnetic fields, there is a non-abelian symmetry replacing it.

The BdG equations 
  are  invariant under the $t$-independent  
  {\it gauge} transform
\begin{equation}\label{gauge-transf'} 
  T^{\rm gauge}_\chi : 	(\g, \al, a)\ra (e^{i\chi }\g e^{-i\chi } , e^{i\chi }  \al e^{i\chi }, a + \nabla\chi),
\end{equation}
where $\chi\in H^2(\R^d, \R)$.  (This defines a natural equivalence relation one should keep in mind.) 
To preserve the Coulomb gauge, we could take $\chi$ {\it linear} in $x$.

Thus, we define the {\it gauge (magnetically) translationally (MT) invariant} states as states  whose translations stay in the same gauge-equivalence class, i.e. which are invariant under the transformations  
\begin{equation*} 
T_{ s}: (  \g, \al, a) \ra (T^{\rm gauge}_{\chi_s})^{-1}T^{\rm transl}_{ s}  (  \g, \al, a),\end{equation*}  
for any $s\in \R^d$, with $d=2, 3$, and for some (linear) functions $\chi_s(x)$. Here $T^{\rm transl}_{ s},\ s\in \R^d,$ is the group of translations. 
We require  $T_{ s}$ to be a projective group representation of $\R^d$. Then, the function $\chi_s(x)$ (of $x$ and $s$) satisfies the {\it co-cycle relation} 
\begin{equation}\label{co-cycle'}
    \chi_{s+t}(x) -\chi_s (x+t) - \chi_t(x)  =\frac 12 b (s, t),
\end{equation} 
 $\forall s, t\in \R^d$, where $b(s, t)$ is a constant 
  two-form on $\R^d$.\footnote{In this case, $T_{ s}$ is, in general, a non-abelian group with the generators which are the components of the operator-vector  $i(-i\n -\frac 12 b\wedge x)$.} Two-forms on $\R^d$, $d=2, 3$, are gauge equivalent to the 2-form $b\cdot (s\wedge t)$, where $b$ is a constant vector if  $d=3$ and a scalar if $d=2$.\footnote{$b\cdot (t\wedge s)$ is the flux of the constant vector field $b$ through the area $t\wedge s$ spanned by the vectors $s$ and $t$.} 
The function $\chi_s(x):= \frac 12 b\cdot (x\wedge s)$ 
  clearly solves the equation \eqref{co-cycle'}. (This fixes a special - symmetric - gauge.)  
  The two-form  $b(s, t)$, or vector/scalar $b$, is identified with a {\it constant external magnetic field}.

\bigskip

This extends the 
 translational symmetry to the  $\curl a\neq 0$ regime.
For $\curl a\neq 0$, gauge-translationally invariant states yield the simplest solutions and  candidates for the ground state for the BdG system.

\bigskip

\paragraph{\bf Ground state.} 
As was alluded to above,  typically, the ground state (GS) has the maximal symmetry. Hence, 
depending on the magnetic field $b$, one expects that:  

\medskip

GS is {\it translationally invariant} for $b= 0$, 

\medskip

GS is  {\it magnetically translationally invariant} for $b\ne 0$.

\medskip

\noindent Candidates for the ground (or equilibrium Gibbs) 
state are:

\smallskip

\begin{enumerate}
\item {\it Normal states}:  $(\g, \al, a)$, with $\al=0$ ($\Rightarrow \g$ is `Gibbs state'). 

\item {\it Superconducting states}:  $(\g, \al, a)$, with $\al \ne 0$ and $a = 0$ (`Meissner states').

\end{enumerate}

\begin{thm}\label{HHSS} 
 For  $b=0$,  
 there is a %unique 
 superconducting, normal,  {\rm translationally invariant} solution. \end{thm}

\begin{thm} \label{CS} %[CS]
  For   $b\neq 0$,  
the MT-invariance implies the normality  $(\al=0)$.
 \end{thm} 
%\smallskip

\begin{cor}    For  $b\neq 0$, the superconductivity  ($\al \ne 0$) implies the symmetry breaking.\end{cor}

An addendum to the maximal symmetry paradigm formulated above could be stated as:  
 in the 
ground state, whenever the maximal symmetry  is broken, it is broken in the most minimal 
 way. For the simplest and most important 
  case of the translationally invariant systems, breaking of the translational invariance leads to formation of crystalline structures with lattice symmetries.  %(crystals, for short)

We are far from being able to prove 
that this happens in reasonable models. However, the natural problem at one level below - the existence and stability of such structures - is possibly approachable.  
 Note that the stability would imply that a crystalline solution 
  is a local minimizer (but not necessarily minimizing locally) of a `renormalized' energy. Though it does not give the global minimizing property, one might say that it is the next best thing.

\medskip

\paragraph{\bf Vortex lattices.}

 Physical experiments show (see below) that most of superconductors in magnetic fields have, in their lowest (locally) energy states, lattice symmetry, yielding minimal symmetry breaking as discussed above. 
Here we define and discuss such states.

 For $b\neq 0$, we define states, which we call 
  the vortex lattice states (or just {\it vortex lattices}), as 

\medskip

\begin{itemize} 
\item 
{\it Vortex lattice}: 
 $(\g, \al, a)$ s.t.  $T^{\rm transl}_{ s} (\g, \al, a) =  T^{\rm gauge}_{\chi_s} (\g, \al, a)$, 
  $\forall s \in \lat$ (some lattice in $\R^2$), with $\chi_s : \lat\times\R^2 \to \R$, and 
  {$\alpha \not= 0$}.  
\end{itemize}

\medskip

The fact that $T^{\rm transl}_{ s}$ is a group representation implies that 
 the map $\chi_s$ satisfies the {\it co-cycle relation} (see \eqref{co-cycle'}): 
\begin{equation}\label{co-cycle}
    \chi_{s+t}(x) -\chi_s (x+t) - \chi_t(x) \in 2\pi \Z,\ \forall s, t\in \lat.
\end{equation}

\noindent Co-cycle relation \eqref{co-cycle} implies that the {\it magnetic flux is quantized}:
 
$$\frac{1}{2\pi} \int_{\Om^{\lat}} \curl a  \in \Z.$$ 
 Here $\Om^\lat$ is a fundamental cell of $\lat$ and  the left-hand side is the 1st Chern number.    The latter can be expressed directly in terms of $\chi$.

%--------------------------- 8 -----------------------------

\bigskip
 
 The existence result for such solutions is recorded in the following 
\begin{thm} \label{thm:BdGExistence}For the BdG system without  the self-interaction term: 

\medskip

(i)  
$\forall n, T>0$ and $\lat$\ 
there is a static  
 solution $u_{T n  \lat}:=(\g, \al, a)$  satisfying 
\begin{align}
	& u_{T n  \lat}\ \text{ is $\lat$-equivariant: }\ T^{\rm transl}_{ s} u_{T n  \lat} = T^{\rm gauge}_{\chi_s} u_{T n \lat }, \forall s\in \lat,\\
&\text{the 1st Chern number is $n$: }\ \int_{\Om^{\lat}}\curl a =2\pi n,\end{align}

\qquad $u_{n T \lat}$ minimizes the {\it free energy}\footnote{For the definition of the free energy see Appendix \ref{sec:BdGsyst}.} $F_T=E-TS-\mu N$ on $\Om^\lat$ for   
$c_1=n$;

\bigskip

(ii) For the pair potential $v\le 0, v\not\equiv 0$ and $T$ and $b$ sufficiently small,  
$u_{T n \lat}$ is a {\it vortex lattice} (i.e.  {$\alpha \not= 0$});

\bigskip
(iii)  For $n>1$, there is a  {\it finer lattice}, $\lat' \supset \lat$
 for which $u_{T n \lat}=u_{T 1 \lat'}$, i.e. 
$u_{T n \lat}$  is $\lat'$-equivariant with $c_1=1$.    
\end{thm}

\paragraph{\it Open problem:} Show that for superconductors of Type II and for $n=1$, the vortex lattice solution is stable. (For discussion of Type I and II superconductors and the notion of stability in the context of the BdG, see \cite{CS1}. These notions as well as the notion of self-duality still need to be elucidated in the BdG theory.)

On the first step, one could address  stability w.r. to symmetry preserving (periodic) perturbations is usually accessible. Showing that a crystalline solution is stable under general perturbations (deforming the lattice in various ways) is a rather subtle, highly non-trivial matter.

\bigskip

%--------------------------- 5 -----------------------------

\section{Ginzburg-Landau equations}

In the leading approximation (close to the critical temperature and after {\it `integrating out'} $\g$ and the relative coordinate $x-y$ of $\al(x, y)$), % and setting $V=0$, % and the relative coordinate $x-y$ in $\al(x, y)$), 
the {\it BdG system} leads to the {\it time-dependent Ginzburg-Landau equations} %Cooper pair WF $\al(x, y)\ra \psi(\frac12(x+y))$, where $\frac12(x+y)$ is  the Cooper pair center-of-mass coordinate 
\begin{equation*}
\bs
   %\g(\partial_{t}+i\phi)
  \g \partial_{t}  \psi &= \Delta_a \psi   + \kappa^2 (1-|\psi|^2) \psi, \\
   %\s(\partial_t a -\nabla\phi)
 \nu \partial_t a  &= -\curl^2 a + \Im ( \bar{\psi} \nabla_a \psi ).
\end{split}
\end{equation*}
%These are
%in the gauge $\phi_{\rm electr}=0$.
  Here $\Re \g, \Re \nu \geq 0$ are constants arising in the approximation. In the application to superconductivity,\footnote{Besides describing equilibrium states of {\it superconductors} (mesoscopically), the static GLE describe also the (static) {\it $U(1)$ Yang-Mills-Higgs model} of particle physics (a part of {\it Weinberg-Salam model of electro-weak interactions}/a standard model). In particle physics, $\psi$ and $a$ are the Higgs and $U(1)$ gauge
(electro-magnetic) fields, respectively.}

\medskip

\qquad {$|\psi |^2$} is the density of 
 superconducting electrons; \\

\qquad $a : \R^{d+1} \to \R^d$ is the magnetic potential;\\

\qquad $\Im ( \bar{\psi} \nabla_a \psi )$ is the superconducting current.

\bigskip
\noindent The second GLE equation comes from two Maxwell equations (Amp\`{e}re's and Faraday's laws). 

\bigskip

\paragraph{\bf Vortex lattices.}As with the BdG and HFB systems, the major open problem is the  symmetry of the ground state. Experiments show that the ground state of a superconductor of Type II in a constant magnetic field is in the form of  
the hexagonal vortex lattice (see Figure 1). %\ref{fig:AL}).  

%--------------------------- 15 picture -----------------------------
%\section{Vortex Lattice. Experiment}

\begin{figure}[h]\label{fig:AL}
\begin{center}

\includegraphics[height=5.0cm]{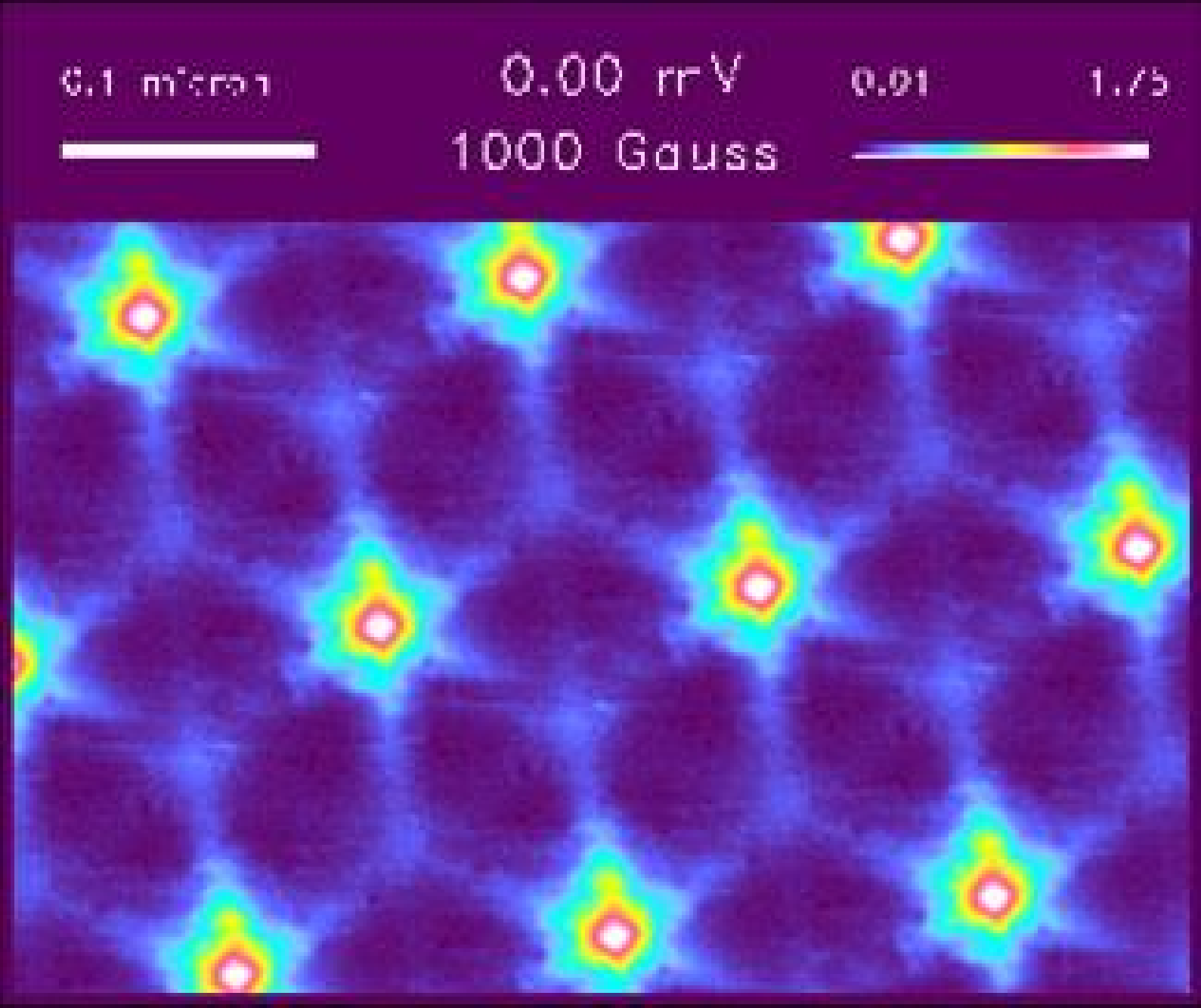}
%%%%%%% the picture can be made smaller/bigger by changing
%%%%%%% height=4.5cm in the command above
\caption{Experimental picture of the Abrikosov lattice obtained using a tunnelling microscope. Different colours
signify different densities of super-conducting electrons.
Theoretical description of this experiment is given solving the BdG system based on the coarse-scale approximation given by the  
Ginzburg-Landau system.}
\end{center}
\end{figure}

For the GLE, the vortex lattice is defined as a {\it static} solution equivariant under the lattice translations in the sense that 
 \[T^{\rm transl}_{ s} (\psi, a) =  T^{\rm gauge}_{\chi_s} (\psi, a),\] for every
  $ s \in \lat$ (some lattice in $\R^2$), 
with $\chi_s : \lat\times\R^2 \to \R$ (satisfying \eqref{co-cycle}).

%--------------------------- 15  -----------------------------

%\section{GLE: vortex lattices}
 
A  vortex lattice solution is formed by {\it magnetic vortices} ({\it localized finite energy} solutions of a fixed degree, see Fig. 2), arranged in a (mesoscopic) lattice $\lat$.  
\begin{figure}[h]

\begin{center}
 \qquad   \qquad   
 
  \includegraphics[height=3.0cm]{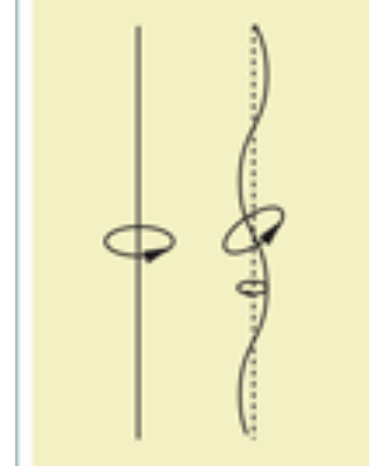} \qquad   \qquad  \text{    }  
  \includegraphics[height=3.0cm]{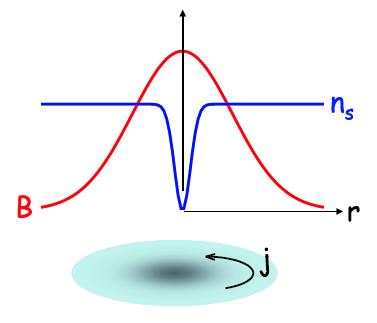}
%%%%%%% the picture can be made smaller/bigger by changing
%%%%%%% height=4.5cm in the command above
%\end{figure}
\caption{The left figure is a 2-dimensional
 section of a picture of a line vortex center. The figure on the right shows density of the
 super-conducting electrons $|\psi|^2=n_s$, the magnitude of the magnetic field and the circulation of the super-conducting current.}

\end{center}
\end{figure}

As for the BdG system,  
 we are far from being able to prove that the ground state of the GLE is given by a vortex lattice. One step below it is the problem of existence and 
 {\it stability} of vortex lattice solutions.  

The {\it existence} of vortex lattice solutions is well known by now.  
 Thus, we are left with the problem: 

\medskip

\begin{itemize}

\item Stability of vortex lattice solutions. \end{itemize}

\medskip

%\medskip

 Since the  
VL {\it are not localized}, the {\it stability is a delicate matter}.

\smallskip

\begin{thm} \label{thm:GLvortlattstab} The vortex lattices are stable under lattice-periodic and local 
perturbations of the same parity as the vortex lattice.\end{thm} 

\bigskip

\noindent {\it Open problem}:  Show stability/instability under more general lattice deformations.

%--------------------------- 15  -----------------------------

\section{Summary}

\begin{itemize}
\item We reviewed some basic properties of the {\it Schr\"odinger equation} which encodes {\it all the information} about quantum systems\footnote{Extraction of the physical information from the solutions is done according to a fixed quantum mechanical procedure. % going back to J. von Neumann.
 The theoretical results agree to remarkable precision with experimental ones, while the physical theory connecting the two, initiated by J. von Neumann soon after advent of Quantum Mechanics, remains largely under construction.}. 
\bigskip

\item  While we learned much about 
  the general structure\footnote{Such as the structure of the spectrum, scattering theory, theory of resonances.} of this equation, our understanding  
of  {\it specific quantum systems}   with the number of particles $\ge 3$ is spotty  and the progress, with some notable exceptions like the stability of matter and the derivation of mean-field type equations, is very slow.  

\bigskip

\item  However, there is one important direction where 
 robust progress is possible -- {\it `effective' 
 equations} for quantum systems of large number of identical particles (quantum gases), 
  especially, those describing quantum fluids (i.e. quantum gases exhibiting some quantum behaviour at the macroscopic scale) such as 
  {\it superconductors, superfluids and Bose-Einstein condensates -- the HFB and BdG equations}. 
  
\medskip

\bigskip

\item  We (a) formulated 
 some key mathematical problems related to  the HFB and BdG equations, 
 (b) described the key stationary solutions of BdG equations, the competitors for the ground/Gibbs state:  {\it normal, superconducting and mixed (or intermediate) states}, and 
(c) presented an important class of the mixed states -- the {\it vortex lattices} -- demonstrating the symmetry breaking.

\end{itemize}

\medskip

%----------------------- 5 -------------------
\section{Remarks on literature}\label{sec:liter}

{\it $n-$particle scattering}. Theorem \ref{thm:AC} is due to {A.~Soffer and I.M.~Sigal ($\mu>1$), and J.~Derezi\'nski ($\sqrt3-1< \mu < 1$)} (\cite{Der}, see also \cite{SigSof1}). The proofs use important earlier results and ideas 
 of P.~Deift and B.~Simon, V.~Enss, C.~G\'erard, G.M.~Graf, 
E.~Mourre, D.~Yafaev (\cite{DS, Enss, FrHe, GGM1,  Ger1, Gra, SigSof3, SigSof4, Skib1, Skib2, Tam, Va1, Va2, Yaf}), see \cite{DerGer1, GerLa,   HunSig2} for books and  a review. 
 
 \bigskip
 
For some of the extensions, see \cite{APSS, HeMoSkib, HeSk, Skib1, Skib2, Skib4, SigSof}).   
 Many ideas and techniques from this field 
 successfully entered  
 Quantum Electrodynamics (QED), see below,  
  nonlinear evolution equations (\cite{BlSof1,  BlSof2, Cucc, CuccMa, DonSchSof1, DonSchSof2, LindSof2, LindSof3, LindSof3, RodSchSof, SchSofSt, Sof}), wave propagation (\cite{DeBievHisSig, DeBievPra1, DeBievPra2}) and 
the energy transfer in the non-autonomous Schr\"odinger equations  
 (\cite{Bamb, BambGrMa, BambGrMaRob, BambLanMon, BerMas, Mas1, Mas2, MasRob, Mon}). 
For an important parallel development see \cite{Va1, Va2}.

\bigskip

{\it NR QED,   
Radiation}. 
Theorem \ref{thm:rad} was proven by V.~Bach, J.~Fr\"ohlich and I.M.~Sigal (\cite{BFS1, BFS2, BFS3}). Many important extensions and improvements were obtained in 
 \cite{AlFau, AlFauGui,  
 ArHirHiros, BFP, BarFauGui, BalFauFrSchu, CaHai, CorrFalOlRoug, DFP1, DFP3, FGSig1, GrHas, GGM2, GuMaMo, Ger1, GrHas, Hai, HaiHirSp, HaiSei1, HaiVouVu, HaHe1,   HaHe2,    HaHe3,   HaHe5,       HaSie, Hiro1, Hiro2, LiLo1, LiLo2, Mo3, TeWa}, to mention some of more recent results, see e.g \cite{Ar2, Gr2, Sig3, Sig5, GS, Sp4} for reviews and book presentations.

The main ingredient in the proof of Theorem \ref{thm:rad} 
 is the spectral renormalization group introduced in  \cite{BFS1, BFS2}.  
 A different approach was developed by V.~Bach, M.~Balestros, J.~Fr\"ohlich, A.~Pizzo, which the authors call the multiscale or Pizzo method (\cite{BachBalInPi, BachBalKoMen, BachBalMen, BachBalPi}).

The infrared (IR) problem was addressed by Th.~Chen, J.~Fr\"ohlich 
 and A.~Pizzo (\cite{ChF, ChFP1, ChFP2}).
 For the renormalization of  the electron mass see  \cite{Ch, BCFS}.

Ideas and techniques from  the NR QED were extended to the Nelson and spin-boson models of the condensed matter physics\footnote{These models have a form similar to the NR QED, with photons replaced by (acoustic) phonons (quantized (longitudinal) oscillations of the underlying medium) and the interaction somewhat modified. The main problems for these models are exactly the same as those formulated in Section \ref{sec:QED}. The main difference is that these models do not have the gauge invariance which allows to lower the IR singularity in the QED Hamiltonian.} in  
 \cite{BaDecPiz, BalDeckHan, BalDeckHan2,  BalDeckFauHan,  Mo2, GeoRas,  DamMo,   HaHe4, HaHinSie, LorMinSp, Skib3, Sp1, Sp3}. 

\bigskip

{\it NR QED, Asymptotic completeness}.  Theorem \ref{thm:Rayl} is due to J.~Faupin and I.M.~Sigal and W.~De Roeck, M.~Griesemer and A.~Kupiainen (\cite{FaupSig, DeRoGriKu}). 
These works used ideas and results of M.~H\"ubner-H.~Spohn, J.~Derezi\'nski-C.~G\'erard, J.~Fr\"ohlich-M.~Griesemer-B.~Schlein (\cite{DerGer2, FrGrSchl1, FrGrSchl2, FrGrSchl3, FrGrSchl4, HuSp, Sp2}) as well as those from the $n$-particle scattering theory discussed above. For further developments 
 \cite{DybMo, DybPi1, DybPi2, DybPi3, BFP, GrZ} and review \cite{Sig5}. 
 
The finiteness of mean number of photons for the  spin-boson model  was proven by W.~De~Roeck and A.~Kupiainen (\cite{DeRoKu}) 
\bigskip

{\it The HFB and BdG systems.}  Stationary versions of these systems  (written in eigenfunction expansion representations) are used extensively in the physics literature.  
 For the HFB  system, one inserts  the delta-function potentials and in the BdG case one sets $a=0$.  
The full, time-dependent systems in the general form as they appear in this paper   were  written out and formally derived in \cite{BBCFS} and \cite{CS1, BenSokSolov}, respectively.

Clearly, the HFB and BdG systems generalize the Hartree and Hartree-Fock equations. For the relation between the Hartree and Hartree-Fock approximations, on one hand, and quasi-free (Wick) states, on the other, see \cite{BenPorSchl} and Appendix E of \cite{CS1}.

For a rigorous derivation of effective equations similar to the HFB and BdG systems, see the books \cite{LSSY, BenPorSchl}, reviews \cite{Lew2, Nap, Roug}  and some recent 
 articles \cite{BenNamPorSchlSeir2, CorrLundRoug1, ChrHaiNam, DecFrPicPiz1, DecFrPicPiz2, DeuSeir, GrillMach2, GrillMach3, NamNap}.
The rigorous theory started (from very different perspectives) with works of K. Hepp, E. H. Lieb and B. Simon and P.-L. Lions \cite{He, LS1, LS2, Lio} (see also \cite{GinVelo1, GinVelo2} for early follow up work). 
%\bigskip

For the {\it HFB system, the static, homogeneous case} (i.e. $V=0$ and $\g$ and $\s$ are translation-invariant) 
  was treated rigorously by   
M.~Napi{\'o}rkowski, R.~Reuvers and J.P.~Solovej, see \cite{NapReuvSolov1, NapReuvSolov2}.  
 In the {\it general case}, general properties and the well-posedness were established in \cite{BBCFS}. 
 
\medskip

The {\it BdG system without the (dynamic) electromagnetic field, i.e. with} $a=0$. See \cite{HaiSei2} for an excellent review. 
Theorem \ref{HHSS} was proven by Ch.~Hainzl, E. Hamza, R.~Seiringer and J.P.~Solovej   (\cite{HHSS}). The Cauchy problem was investigated  by N.~Benedikter, J.~Sok and J.P.~Solovej (\cite{BenSokSolov}). 

 For the {\it full BdG system}, general properties of the system and classification and properties of the ground states (and more generally, static solutions) were established by I. Chenn and I.M. Sigal (\cite{CS1}). In particular, Theorem \ref{CS} was proven in \cite{CS1}.  An asymptotic behaviour of critical temperature in weak magnetic fields was established in \cite{FrHaiLang, DeuHaiScha}.

For $\al=0$, the BdG system leads to the Hartree-Fock equation coupled to the Maxwell equations.  Closely related to the Hartree-Fock equation is the important and widely used    {\it Kohn-Sham equation}, the main tool in the {\it density functional theory (DFT)}, see \cite{AnCan, CanM, CLS, CS0, ELu2, ELu3, ELu, ELY, PrNord, PusS}  
 and references therein for the former  and \cite{Lew3,  LewinLiebSeir, LewLiebSeir2} for the latter.

In a remarkable work, R.~Frank, Ch.~Hainzl,  R.~Seiringer, J.P.~Solovej  (\cite{FHSS}) have shown that, for non-dynamical magnetic fields, the nanoscopic approximation of the BdG system is given by the Ginzburg-Landau one (see also \cite{FHSS2, FHSS3, CFHS}).

\medskip

{\it Vortex lattices}. {Existence of vortex lattices for the BdG equations} was proven by   I.~Chenn and I.M.~Sigal (\cite{CS1}).

\smallskip

For the existence results for the {\it Ginzburg-Landau system},  
 see review \cite{Sig2}. 
Stability of vortex lattices for the Ginzburg-Landau system under lattice-periodic and local perturbations was proven by
I. M.~Sigal and  T.~Tzaneteas (\cite{ST1, ST2}).

\medskip

 Important  
  results on asymptotic behaviour of solutions, for $\kappa \to\infty$ and applied magnetic fields, $h$, satisfying $h\le \frac{1}{2}\ln \kappa$+const (the London limit), were obtained in \cite{ASand}.  
  Further extensions to the Ginzburg-Landau equations for anisotropic and high temperature superconductors in the $\kappa \to\infty$ regime can be found in \cite{ABS1, ABS2}.

For 
 the $\kappa\ra \infty$ (the semi-classical) 
  regime in the Ginzburg-Landau equations 
and the linear eigenvalue problem related
to the second critical magnetic field, see the books  
\cite{SS}\footnote{For the precursor of the development described in this book, see \cite{BBH}.} and \cite{FournHelffer}, respectively, and see  \cite{AftSe, AlHe, AlHePa, CorrRoug1, CorrRoug2, CorrRoug3, FouHe1, FouHe2, FouHePe,  FournKach1, FournKach2}, for some additional and more recent results. 
 
\medskip

There are many similarities between  the Ginzburg-Landau and {\it Gross-Pitaevskii} equations and more specifically between the phenomena of superconductivity and the Bose-Einstein condensations (BEC) described by these equations, respectively. The key results for  the BEC 
vortices in the Gross-Petaevski  equation are due to  A.~Aftalion, R.~Jerrard, M.~Correggi, N. Rougerie,  J.~Yngvasson et al, 
  see the book \cite{Aft} and the more recent papers \cite{AftBlJe,  AftMaWe, AftNoSou, AftSa, AftSou, CorrRougYng, CorrDubLundRoug}.

 For recent work on vortices in the somewhat similar {\it Landau-Lifshitz}-type equations and the Weinberg-Salam model of electro-weak interactions (the $U(2)$ Yang-Mills-Higgs system), see \cite{GuWa, LiMe, GarS}.

The HFB, BdG,  GL and GP equations are obtained  from the original many-body Schr\"odinger equation by `integrating out' some degrees of freedom corresponding to finer length scales (the GL and GP equations) or faster dynamics (the HFB and BdG equations). 
 Such equations are called the effective equations and the dynamics described by them, the effective dynamics. One can continue further integrating out degrees of freedom   
  to obtain even coarser (say, nanoscopic) effective equations, see e.g. \cite{BrJer, CollJer, CLaLe, CCoSc, CSc,  CorrRougYng, CorrDubLundRoug, DFP2, ELu2, ELu3, ELu, ELY, FeffWein3, FeffLThWein2, FeffLThWein3,  FrGJS, GangS1, GangS2, Jer, JerSm, JFrGS, Holm, HMZw, HPZw, HZw, HZw2, WaLuWein}.

For effective equations approach to other quantum-mechanical problems, see \cite{Lew1, Lew4,  Li2, Li4, SimK, Te} for reviews and books   and \cite{AnaLew, AnaSig, BrMa, FrKnSchSo, FrLem, FrNamVan1, FrNamVan2, LiSei, MaSo, Nier, Pan, PST1, PST2}, 
for some recent work.   
Of important aspects of the quantum many-body problem lying further a field  
 and not discussed above,  
  we mention the Coulomb systems, positive temperatures, topological techniques, 
  random Schr\"odinger operators 
  and perturbation theory. For some general reviews and books in these areas, see \cite{AtJoPil, Avr, AFG, CFKS, DiKiKleKlo, DSS, Feff1, Feff2, Fr3, Fr4, GJ, His, JP, Kir, LeBL, Li1, Li3, LiSei, SimH, SimK}.

\appendix

%%%%%%%%%%%%%%%%%%%
\section{The NR QED Hamiltonian}\label{sec:QED-ham}

In this appendix, we define the quantized vector potential and  the Hamiltonian of the quantized electromagnetic field entering the Hamiltonian \eqref{Hqed}.

First, we mention that  the state space 
$ \cH_f$ of photons is  the bosonic Fock space, $\cF:=\oplus_{n=0}^\infty \otimes_s^n \fh$, where $\otimes_s^0 \fh:=\C$, based on the one-photon space $\fh:=L^2(\R^3, \C^2)$   ($\otimes_s^n$ stands for the symmetrized tensor product of $n$ factors,
 $\C^2$ accounts for the photon polarization). 
 
 There is a distinguished pair of   unbounded, operator-valued distributions, $a_\lambda(k)$ and $a_\lambda^*(k)$, acting on $\cF$, called the annihilation and creation operators, which generate the operator algebra on $\cF$.\footnote{$k$ is the photon wave vector and $\lambda$ is the photon polarization.} They
  are defined as follows.
With each function $f \in \fh$, one associates the %\emph{creation}  and \emph{annihilation operators}
operators $a(f)$ and $a^*(f)$ defined, for $u\in \otimes_s^n\fh,$ as
\begin{equation*} %\label{cr-annih-ops}
a^*(f)  : u\ra \sqrt{n+1}  f\otimes_s u\ \quad   \mbox{and}\  \quad  a(f)  : u\ra \sqrt{n}  \lan f, u\ran_\fh, %\tag{I.1}
\end{equation*}
with %$\lan f, u\ran_\fh:=\int \overline{f(k)} u(k, k_1, \ldots, k_{n-1}) \, d k$, for phonons, and 
 $\lan f, u\ran_\fh :=\sum_{\lambda = 1 , 2} \int  \, d k \overline{f(k, \lam)} u_{n}(k, \lam, k_1, \lambda_1,$ $ \ldots, k_{n-1}, \lambda_{n-1})$. %for photons. 
\DETAILS{They are unbounded, densely defined operators of $\G(\fh)$, adjoint of each other (with respect to the natural scalar product in $\cF$) and satisfy the \emph{canonical commutation relations} (CCR):
\begin{equation*}
\big[ a^{\#}(f) , a^{\#}(g) \big] = 0 , \qquad \big[ a(f) , a^*(g) \big] = \lan f, g\ran ,
\end{equation*}
where $a^{\#}= a$ or $a^*$.} 
%(Operators $a(f)$ and $a^*(f)$ are adjoint of each other.) 
($a^*(f)=(a(f))^*$.) Since $a(f)$ is anti-linear and $a^*(f)$ is linear in $f$, we can write formally %, for photons,
\begin{equation}\label{a-a*}
a(f) = \sum_{\lambda = 1 , 2} \int \overline{f(k, \lam)} a_\lambda(k) \, d k , \qquad a^*(f) = \sum_{\lambda = 1 , 2} \int  f(k, \lam) a_\lambda^*(k) d k 
\end{equation}
and consider $a_\lambda(k)$ and $a_\lambda^*(k)$ as formal objects with certain commutation properties.

  The operators $A_\xi$ and $\hf$,  describing the quantized electromagnetic field and its dynamics, respectively, are  given by
\begin{align} \label{Aem}
&A_\xi(y)=\sum_{\lambda=1,2} \int\frac{ \xi (k) d k}{\sqrt{2 \omega ( k )}}  \varepsilon_\lambda (k) \big( e^{i k \cdot y} a_\lambda (k) + e^{- i k \cdot y} a_\lambda^* (k) \big),\\
 \label{Hf}
&\hf \ = \  \sum_{\lambda=1,2} \int d^3 k \; \om(k) a_\lambda^*(k)  a_\lambda(k).
\end{align}
Here, $\omega ( k ) = \vert k \vert$ denotes the (Einstein) photon dispersion relation (the energy as a function of the  wave vector). 

To give rigorous meaning to expressions like \eqref{Aem} and \eqref{Hf}, we can express them in terms of $a(f_j)$ and $a^*(f_j)$ for some orthonormal basis $\{f_j\}$ in $\fh$.

%--------------------------- 1 -----------------------------

\section{Hartree-\-Fock-\-Bogo\-lubov equations}\label{sec:HFBsyst}
 
For the pair interaction potential  $v$,  
the {\it HFB} equations are 
 \begin{align}
    i\partial_{t}\phi & = %h_{\lam}(t)\phi_t %(h+2\lam\rho_{\gamma_t})
    h\phi +\lam|\phi|^{2}\phi +2\lam\rho_{\gamma }\phi +\lam\bar{\phi }\rho_{\alpha },
    \tag{GP}\\ %\label{HFB-NQDdelt1}\\
    i\partial_{t}\gamma  & =[h_{\gamma , \phi },\gamma ]_- + \lam [w_{\al , \phi} , \alpha ]_-, 
    \tag{HF}\\ % \label{HFB-NQDdelt-2}\\
    i\partial_{t}\alpha & =[h_{\gamma , \phi} ,\alpha ]_{+}
    + \lam [w_{\al , \phi} , \gamma ]_{+}     +\lam w_{\al , \phi},    \tag{Coh} % \label{HFB-NQDdelt-3}
\end{align}
where %$[A,B]_+=AB^T+BA^T$ with $A^T=\bar A^*$, %$A^T(x,y)=A(y,x)$, 
$h$ is a one-particle Schr\"odinger operator,   $[A,B]_\pm$  are defined after \eqref{Amp-Maxw-eq} %\footnote{\ms{$[A,B]_-=A B^* - B A^*$, $[A,B]_+=A\bar B^* \pm B\bar A^*$, with $\bar A:=\cC A\cC$, with $\cC$ being the complex conjugation.}}
  and  %$[A,B]_+=AB^T + BA^T$, with $A^T=\bar A^*$, %$\rho_\g(x, t):= \g (x; x, t)$, 
\begin{align}  \notag %\label{eq-nmwh-notations-0}
& \rho_\sigma(x, t):= \sigma (x;x, t), \quad  w_{\al , \phi}:=\rho_{\alpha } +\phi^2,\\ 
    &  h_{\gamma , \phi }:=h+2\lam (\rho_{\gamma }+ |\phi |^{2}) +ex(\g)\,.
\end{align} 

\medskip

The last term on the r.h.s. of Eq. (Coh), $\lam w_{\al , \phi} := \lam (\rho_{\alpha } +\phi^2 )$, is the quantum depletion term.  
Because of it the system has no solutions of the form $(\phi, 0, 0)$ ($100\%$ condensation). %, % and $(0, 

If this term is omitted then the system becomes Hamiltonian and  has solutions of the form $(\phi, 0, 0)$, 
where $\phi$  
satisfies the {\it Gross-Pitaevskii equation}.

%--------------------------- 1 -----------------------------

\section{Bogolubov-de Gennes Equations: Discussion}\label{sec:BdGsyst}
The BdG equations \eqref{BdG-gam}-\eqref{Amp-Maxw-eq} are derived from the 
von~Neumann-Landau equation  \eqref{vN-eq} with Hamiltonian \eqref{Hqed} with $m_j=m$ and \[U(x)=\sum_1^n V (x_i) +\sum_{i<j} v (x_i-x_j).\] 
This Hamiltonian is written in the Coulomb gauge with $A$ being the quantized transverse vector  potential and with %longitudinal part of  the classical vector  potential expressed as the Coulomb potential and
 the electrostatic potential incorporated into $\sum_{i<j} v (x_i-x_j)$.
To obtain the BdG equations \eqref{BdG-gam}-\eqref{Amp-Maxw-eq}, we pass to  the Wick (or quasifree) approximation as in Section \ref{sec:eff-eqs}, but, in addition, use the coherent states in the vector potential. The effective magnetic potential $a$ arises from $A$.

It turns out that  it is natural to organize the operators  $\gamma$ and $\s$ 
 into the self-adjoint operator-matrix (called the generalized one-particle reduced density matrix)
\begin{align} \label{Gam}
	\eta:= \left( \begin{array}{cc} \g & \al \\  \al^* & \one-\bar\g \end{array} \right).
\end{align}
Similarly to $\g$,
one can then show that this  the operator-matrix $\eta$ obeys
 the inequalities\footnote{The quantum fields (the annihilation and creation operators) $\psi(x)$ and $\psi^{*}(x )$, used in Section \ref{sec:eff-eqs}, are operator-valued distributions (cf. Appendix \ref{sec:QED-ham}). If we write them as functionals $\psi(u)=\int \psi(x) u(x) dx$ and $\psi^*(u)=\int \psi^*(x) \bar u(x) dx$ (cf. \eqref{a-a*}), then \eqref{gam-def}-\eqref{al-def} can be rewritten as $\chi(\psi^*(v)\psi(u))=\lan u, \g v\ran$ and  $\chi(\psi(v)\psi(u))=\lan \bar v, \al u\ran$. Letting $f=(f_1, f_2)\in L^2(\R^d)\oplus L^2(\R^d)$ and $A(f):=\psi(f_1)+\psi^*(\bar f_2)$ and using the previous equations, one can easily compute  that $\chi(A^*(g)A(f))=\lan f, \eta g\ran$, which, together with the anti-commutation relation $\{A^*(g), A(f)\}=\lan f, g\ran$, yields $0\le \eta=\eta^* \le 1$.} 
	\[0\le \eta=\eta^* \le 1.\] 
These inequalities imply that 
 the bounded operators $\g$ and $\al$ 
   satisfy the relations 
\begin{equation}  \label{gam-al-prop}  
	0\le \gamma=\g^* \le 1,\  \quad 	 \al^*= \overline\al\ \quad  
	  {\rm\ and\ }\  \quad \alpha \alpha^* \leq \gamma(1-\gamma),
\end{equation}
where  $\overline\gamma :=\cC \g \cC$, with $\cC$, the operation of complex conjugation  
 (see \cite{CS1}, Appendix E, and \cite{BenSokSolov}).

Furthermore, the first two equations, \eqref{BdG-gam}-\eqref{BdG-al}, of the BdG system can be written in the form similar to the HF equation \eqref{HF-eq}:  
 \begin{align}\label{BdG-eq-t}
&% i (\partial_t + i \phi)
i \partial_t  \eta =[H_{a, \eta}, \eta], \quad \text{ where }\  \quad  H_{a, \eta} =  \big(\begin{smallmatrix} h_{a, \g} & v^\sharp \al\\ v^\sharp \bar{ \al} & -\overline{h_{a,  \g}}\end{smallmatrix}\big), \end{align} 
with $h_{a, \g}$ and $v^\sharp \al$ defined in Section \ref{sec:BdGeqs}.

 \bigskip

The stationary BdG equations arise as 
the Euler-Lagrange equations  for the (BCS) free energy functional 
\begin{align}\label{FT-def} F_{T}(\g, \al, a):=E(\g, \al, a) -T S(\g, \al)-\mu N(\g),\end{align}
where $S(\g, \al) = \Tr g(\G)$, with $	g(\lam):= -\lam \ln \lam - (1-\lam)\ln (1-\lam)$, 
 is the entropy, 
 $N(\g):=\tr \g$ is the number of particles, and $E(\g, \al, a)$ is the conserved energy functional   for $\g, \al$ and $a$  time-independent given by 
 \begin{align} \notag 
	E(\g, \al, a) &=\Tr\big((-\Delta_a) \gamma \big) +\frac12\Tr\big((v* \rho_\g) \gamma \big) -\frac12\Tr\big((v^\sharp \g) \gamma \big)\\
 \label{energy} & +\frac{1}{2} \Tr\big( \al^* (v^\sharp \al) \big)  
 +\int  |\curl a |^2. 
\end{align}

Not surprisingly, it turns out that $E(\g, \al, a):=\qf(H_a)$, where  $\qf$ is a Wick (quasi-free) state in question and  $H_a$  is the standard many-body Hamiltonian, %given in \eqref{H}, 
coupled to the vector potential $a$.

%%%%%%%%%%%%%%%%%%%%%%%%%
%%%%%%%%%%%%%%%%%%%%%%%%%%
\end{document}